\documentclass[lettersize,journal]{IEEEtran}
\usepackage{amsmath,amsfonts}
\usepackage{algorithmic}
\usepackage{algorithm}
\usepackage{array}
\usepackage[caption=false,font=normalsize,labelfont=sf,textfont=sf]{subfig}
\usepackage{textcomp}
\usepackage{stfloats}
\usepackage{url}
\usepackage{verbatim}
\usepackage{graphicx}
\usepackage{cite}
\begin{document}

\title{CNN-Assisted Particle Swarm Optimization of a Perturbation-Based Model for Nonlinearity Compensation in Optical Transmission Systems}

\author{Alexey Redyuk, Evgeny Shevelev, Vitaly Danilko, and Mikhail Fedoruk
\thanks{The authors are with Novosibirsk State University, Novosibirsk, 630090, Russia.}}

\markboth{Journal of \LaTeX\ Class Files,~Vol.~14, No.~8, August~2021}%
{Shell \MakeLowercase{\textit{et al.}}: A Sample Article Using IEEEtran.cls for IEEE Journals}

\maketitle

\begin{abstract}
Nonlinear signal distortions are one of the primary factors limiting the capacity and reach of optical transmission systems. Currently, several approaches exist for compensating nonlinear distortions, but for practical implementation, algorithms must be simultaneously accurate, fast, and robust against various interferences. One established approach involves applying perturbation theory methods to the nonlinear Schr\"{o}dinger equation, which enables the determination of the relation between transmitted and received symbols. In most studies, gradient methods are used to find perturbation coefficients by minimizing the mean squared error between symbols. However, the main parameter characterizing the quality of information transmission is the bit error rate.
We propose a modification of the conventional perturbation-based approach for fiber nonlinearity compensation in the form of a two-stage scheme for calculating perturbation coefficients. In the first stage, the coefficients are computed using a convolutional neural network by minimizing the mean squared error. In the second stage, the obtained solution is used as an initial approximation for minimizing the bit error rate using the particle swarm optimization method. In numerical experiments, using the nonlinearity compensation algorithm based on the proposed scheme, we achieved a 0.8~dB gain in the signal-to-noise ratio for a 16QAM 20$\times$100 km link with a channel rate of 267~Gbit/s and demonstrated improved accuracy compared to the single-stage scheme. We estimated computational complexity of the algorithm and demonstrated the relation between its complexity and accuracy. Additionally, we developed a method for learning perturbation coefficients without relying on ideal symbols from the transmitter, instead using the received symbols after hard decision detection.
\end{abstract}


\section{Introduction}
\IEEEPARstart{G}{lobal} data traffic transmitted by optical communication systems continues to grow by tens of percent per year, driving further advancements in existing transmission technologies. However, fiber-optic communication links face significant limitations in increasing capacity and transmission range, primarily due to nonlinear signal distortions \cite{winzer_oe_2018, soman_jo_2021, zhitelev_qe_2016}. To address this, various digital signal processing approaches have been proposed to compensate for optical channel nonlinearity. For practical application, these approaches require algorithms that are highly accurate, fast in implementation, and resistant to interference. One well-known method is the backpropagation technique \cite{ip_jlt_2010}, which models signal propagation in the reverse direction at the receiver. Despite its practical implementation complexity, this method is often used as a benchmark for comparing the effectiveness of other algorithms. Additionally, the use of machine learning techniques to compensate for channel nonlinearity \cite{sidelnikov_jlt_2021, sidelnikov_oe_2018}, jointly compensate for multiple effects \cite{freire_jlt_2024, khan_jlt_2019, hager_ofc_2020}, and optimize transmission parameters \cite{karanov_jlt_2018, zibar_jlt_2020} is also being actively developed.

It is well known that the propagation of a communication optical signal along a multi-span fiber communication link can be described by the nonlinear Schr\"{o}dinger equation (NLSE). One approach to compensating for nonlinear distortions involves applying perturbation theory to the NLSE, which allows for the determination of the relation between transmitted and received symbols \cite{tao_jlt_2011, sorokina_oe_2016, redyuk_jlt_2020}. In recent years, numerous studies have focused on increasing the efficiency and reducing the computational complexity of this method by optimizing perturbation coefficients \cite{barreiro_jlt_2023, kozulin_oc_2021, barreiro_preprint_2024} and employing neural networks \cite{luo_arhiv_2022, ding_jlt_2022, melek_oft_2021, luo_ph_2022}. However, most studies aimed at optimizing this method use gradient schemes that minimize the mean squared error (MSE) between transmitted and equalized received symbols, which can sometimes yield unexpected results \cite{freire_jstqe_2022}. While MSE is a useful metric for gradient calculations, the bit error rate (BER) is a more pertinent parameter for evaluating the quality of information transmission.

In this work, we propose a modification of the existing perturbation-based model for compensating nonlinear distortions using a two-stage scheme for calculating perturbation coefficients. In the first stage, a convolutional neural network (CNN) is employed to calculate the coefficients by minimizing the MSE. In the second stage, the obtain solution serves as an initial approximation for minimizing the BER using the probabilistic particle swarm optimization method (PSO). The performance of the proposed scheme is demonstrated on simulated data for a 20$\times$100~km communication link and a 16QAM signal with a channel rate of 267~Gbit/s. The accuracy of the proposed algorithm is shown to be improved compared to the single-stage approach, and the computational complexity of the algorithm is evaluated.

\section{Perturbation Theory for Nonlinear Distortion Compensation}
Several studies have focused on models for compensating nonlinear signal distortions based on perturbation theory \cite{tao_jlt_2011, redyuk_jlt_2020, sorokina_oe_2016, barreiro_jlt_2023, kozulin_oc_2021, barreiro_preprint_2024, luo_arhiv_2022, ding_jlt_2022, melek_oft_2021, luo_ph_2022}.
These works are founded on the premise that the propagation of optical pulses along a fiber-optic communication link can be described by the NLSE:
\begin{equation}
\frac{\partial A}{\partial z} = \left[-\frac{\alpha}{2} - i\frac{\beta_2}{2}\frac{\partial^2}{\partial t^2} + i\gamma|A|^2\right] A,
\label{nlse}
\end{equation}
where $\alpha$, $\beta_2$, and $\gamma$ are the attenuation, dispersion and Kerr coefficients, respectively. $A(z,t)$ represents the complex field envelopes of the optical signal with initial conditions given by $A(0, t) = \sum_{k} a_{k} f(t - kT)$. Here, $f(t)$ is the carrier pulse profile, $T$ is the symbol interval duration, and $a_k$ are the complex symbols from the used alphabet. By considering nonlinearity as a small parameter and applying perturbation theory, the following relation can be obtained, linking the symbols $a_k$ at the transmitter and the symbols $b_k$ after compensating for all linear effects at the receiver:
\begin{equation}
a_k \approx C_0 b_k + \sum_{m,n=-M}^{M} C_{m,n}b_{k+m}b_{k+n}b^*_{k+m+n} + C_b,
\label{pbm}
\end{equation}
where $M$ is the channel memory parameter, $C_0$, $C_b$, and $C_{m,n}$ are the complex perturbation coefficients, and $b^*$ denotes the complex conjugate of $b$.
The significance of this relation lies in its potential as the basis for an algorithm to compensate for nonlinear distortions. By knowing the perturbation coefficients and receiving a stream of symbols $b_k$ at the receiver, one can use equation (\ref{pbm}) to approximate the transmitted symbols $a_k$.
In simpler cases, the perturbation coefficients have analytical expressions in the form of multidimensional integrals, which depend on the fiber parameters, signal power, and pulse shape.
However, these expressions are often cumbersome and complex for both analytical and numerical calculations.
In most recent studies, the coefficients are computed using machine learning methods or gradient methods that minimize the MSE between the symbols $a_k$ and $\hat b_k$, obtained after nonlinear compensation using perturbation theory (\ref{pbm}).
These approaches require training and test sets of symbols, which can be obtained through simulations, experiments, or the use of pilot sequences.

\begin{figure*}[!t]
\centering
\includegraphics[width=6.9in]{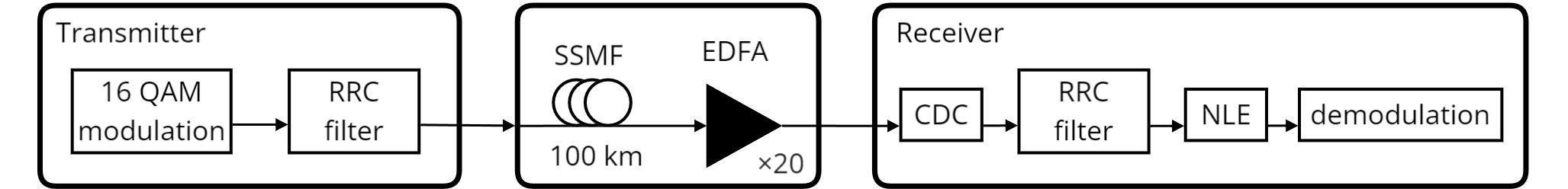}
\label{fig_second_case}
\caption{Schematic of the transmission link under investigation: SSMF -- standard single-mode fiber, EDFA -- erbium-doped fiber amplifier, CDC -- chromatic dispersion compensation, NLE -- nonlinear equalizer.}
\label{fig1}
\end{figure*}

One of the primary characteristics of any method for compensating nonlinear distortions is its performance. However, from a practical perspective, the number of arithmetic operations required to recover a single symbol is equally important. Therefore, developing such a method involves a bi-criteria optimization: enhancing performance while simultaneously reducing computational complexity.
In this work, we evaluate the computational complexity of the method by counting the number of complex multiplications required to compute one symbol according to equation (\ref{pbm}).

\section{Transmission System Model}
Figure \ref{fig1} illustrates the communication link setup used to test the proposed approach.
The model consists of a transmitter, 20 spans of standard single-mode fiber, each 100~km long, erbium-doped fiber amplifiers (EDFA) after each span, and a signal receiver.
At the transmitter, random 16QAM symbols are generated to form a pulse sequence.
The pulses are shaped using a root-raised-cosine (RRC) filter with a roll-off factor of 0.1 and a symbol rate of 66.7~GBaud, corresponding to a bit rate of 267~Gbit/s in a single polarization.
This work studies the propagation of a single spectral channel with one polarization, with the central wavelength of the signal band set at 1550~nm. Noise induced by the EDFA with a noise figure of 4.5~dB, is added to the optical signal after each span as additive white Gaussian noise.
Each EDFA amplifier precisely compensates for the signal loss in one span.

Nonlinear propagation of signals through optical fiber is modeled by the NLSE (\ref{nlse}).
Detailed fiber and signal parameters are listed in Table \ref{parameters}.
The equation was numerically solved using a symmetric split-step Fourier method with a sampling rate of 16 samples per symbol and a logarithmic step size in $z$.
After propagation through the link, chromatic dispersion compensation was performed, followed by matched filtering and downsampling to 1 sample per symbol, along with phase drift correction.
Subsequently, compensation for nonlinear distortions was performed, followed by symbol demodulation, conversion into bits, and calculation of the BER.

\begin{table}[b]
\renewcommand{\arraystretch}{1.3}
\caption{Transmission model parameters}
\label{parameters}
\centering
\begin{tabular}{ll|ll}\hline
Parameter & Value & Parameter & Value \\\hline
Attenuation, $\alpha$ & 0.2 dB/km & COI Wavelength & 1550 nm\\
Dispersion, $\beta_2$ & -21.7 ps$^2$/km & RRC Roll-off & 0.1\\
Nonlinearity, $\gamma$ & 1.4 1/W/km & Symbol Rate & 66.7 GBaud\\
Length, $L$ & 20$\times$100 km & Channel Spacing & 66.7 GHz\\
Noise Figure & 4.5 dB & Channel Data Rate* & 267 Gbit/s\\\hline
\multicolumn{4}{r}{*including FEC overhead}
\end{tabular}
\end{table}

\begin{figure}[!t]
\centering
\includegraphics[width=2.8in]{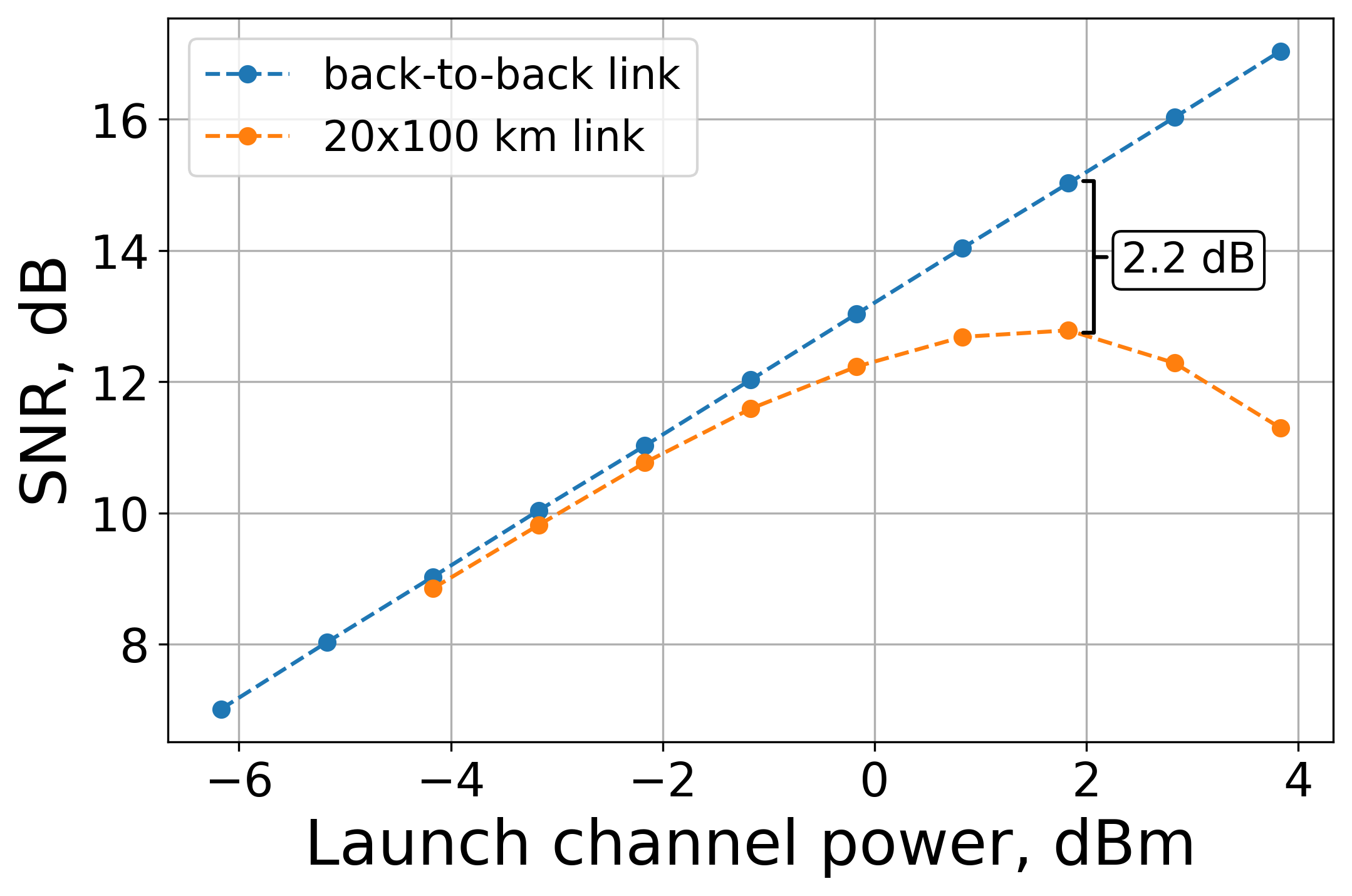}
\caption{Dependence of the signal-to-noise ratio on the input signal power for the ``back-to-back'' configuration and the complete model of the link.}
\label{fig2}
\end{figure}

The relationship between launched signal power and information transmission quality was investigated for this link configuration. The signal-to-noise ratio (SNR) was used as the quality metric, calculated directly from the bit error rate using the formula
$SNR_{\text{16QAM}} = 20 \lg\left[ \sqrt{10} \, \text{erfc}^{-1}\left( 8 \text{BER} / 3 \right)\right],$
where $\text{erfc}^{-1}(x)$ denotes the inverse complementary error function $\text{erfc}(x)$.
Figure \ref{fig2} illustrates the SNR dependency on the input signal power for both the full link model and the ``back-to-back'' configuration without spans, where noise equivalent to that from all amplifiers was added at the receiver. In the full model, perfect compensation of chromatic dispersion, matched filtering, downsampling, and phase offset correction were applied at the receiver.

From Figure \ref{fig2}, it is evident that for signal powers above 0~dBm, the curve corresponding to the full model starts to deviate from the ``back-to-back'' curve. This deviation indicates the increasing influence of nonlinear signal distortions due to higher signal power and the role of nonlinear effects.
To investigate the characteristics of the proposed nonlinear distortion compensation method, we fixed the signal power at 2~dBm, which corresponds to the SNR peak. At this signal power, the BER is 0.02, the SNR is 12.7~dB, and the SNR penalty due to nonlinearity is found to be 2.2~dB. All subsequent analyses were performed on simulated data at this specified signal power of 2~dBm.
For this purpose, independent calculations were used to generate training and testing sets of symbols, each with a length $2^{17}$. 
\begin{figure*}[!b]
\centering
\includegraphics[width=6.4in]{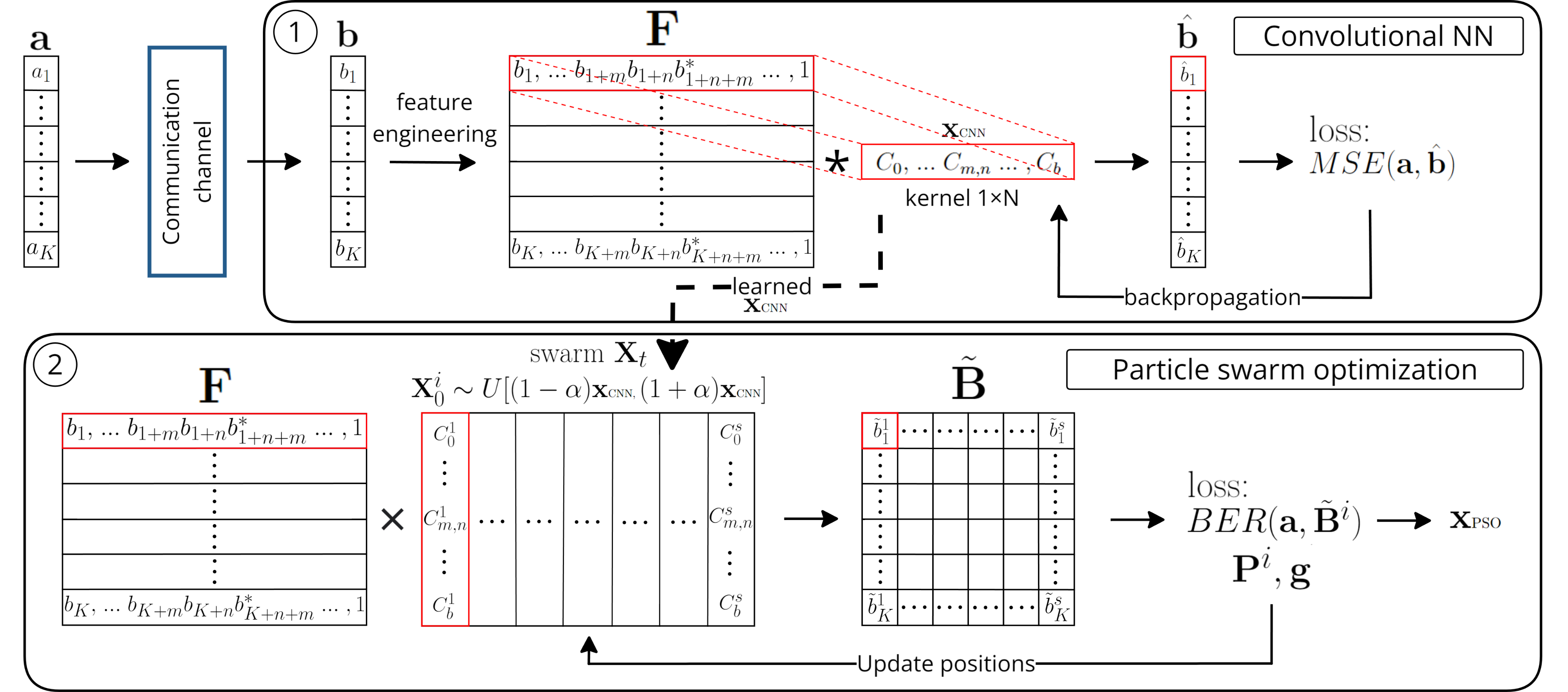}
\caption{Scheme of the proposed two-stage approach for computing perturbation coefficients.}
\label{fig_}
\end{figure*}
The training set was used only to train perturbation coefficients, while the testing set was utilized to compute all metrics, notably the signal-to-noise ratio improvement $\Delta SNR = SNR_{PBM} - SNR_{CDC}$, where $SNR_{CDC}$ and $SNR_{PBM}$ denote the SNR before and after applying the distortion compensation algorithm, respectively.

\section{Two-Stage Scheme for Computing Perturbation Coefficients}
We introduce a novel two-stage scheme for computing perturbation coefficients $C_{m,n}$, $C_0$ and $C_b$. As shown in Figure~\ref{fig_}, in the first stage a convolutional neural network is used to estimate coefficients by minimizing mean squared error between transmitted and equalized symbols. In the second stage, these coefficients are refined using the particle swarm optimization method to minimize bit error rate.

To elaborate on our proposed scheme, we first need to examine the right side of equation (\ref{pbm}). This equation takes the form of a linear equation with respect to the unknown variables $C_0$, $C_{m,n}$, $C_b$ with known coefficients $b_k, ..., b_{k+m}b_{k+n}b^*_{k+m+n}, ..., 1$. These elements are referred to as the features of the symbol $b_k$. Let $\mathbf{F}$ denote the feature matrix, where each row $k$ contains the features of symbol $b_k$ from the entire symbol set.

Our approach starts finding the perturbation coefficients using the rows of the feature matrix $\mathbf{F}$ as input to a single convolutional layer with a learned kernel size of $1 \times N$, which will consist of the coefficients $C_0$, $C_{m,n}$, $C_b$. However, to ensure manageable computational complexity during the implementation of this approach, it is necessary to reduce the number of perturbation coefficients.

\subsection{Feature Engineering}
Using the analytical expression for the coefficients \( C_{m,n} \), it can be shown that three symmetries apply to the indices \( m \) and \( n \): \( C_{m,n} = C_{n,m} = C_{-m,-n} = -C^*_{-m,n} \) \cite{redyuk_jlt_2020, sorokina_oe_2016}.
Incorporating these symmetries into equation (\ref{pbm}) significantly reduces the number of unknown coefficients \( C_{m,n} \), practically by a factor of eight, thus lowering the computational complexity of the method. Taking it into account, the number of complex multiplications required to reconstruct a single symbol using equation (\ref{pbm}) is given by \( C = 9M^2/2 + 19M/2 + 4 \).

\begin{figure}[htbp]
\centering
\includegraphics[width=3.45in]{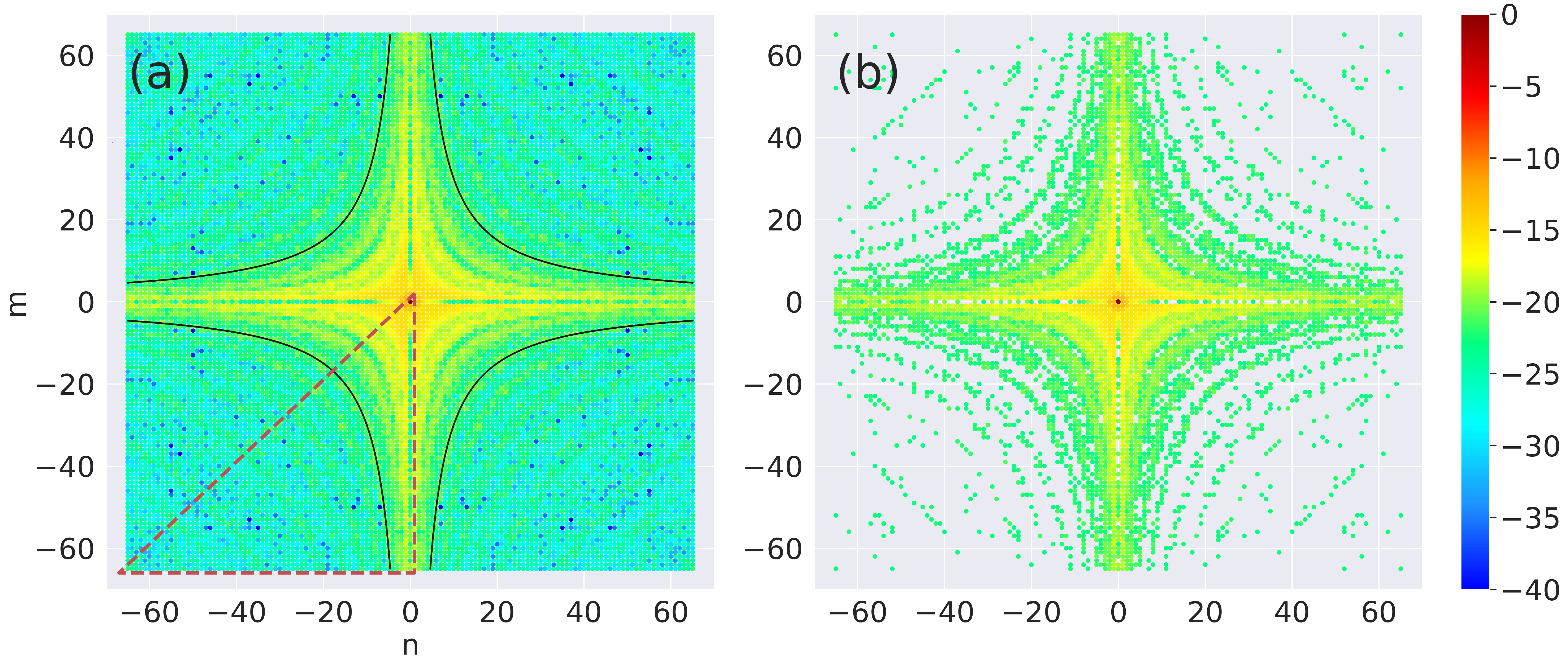}
\caption{Distribution of the magnitude of coefficients \( C_{m,n} \) as a function of indices \( m \) and \( n \) for \( M=65 \), considering the exclusion of coefficients based on the rule \( |mn|<Q \) (a) and the rule \( |C_{m,n}|>Q \) (b).}
\label{fig3}
\end{figure}

Figure \ref{fig3}a represents the magnitude of calculated coefficients depends on indices \( m \) and \( n \). 
Highlighted by the red dashed line is the area of unique coefficients that need to be determined, considering three symmetries.
It is evident that coefficient \( C_{00} \) has the largest value, while the magnitudes of other coefficients decrease as \( |m| \) and \( |n| \) increase.
This observation underpins methods aimed at reducing the computational complexity of the algorithm by discarding coefficients with smaller magnitudes, which minimally impacts the performance.

The first method referred as geometrical pruning involves retaining coefficients that satisfy the condition \( |mn| < Q \), where \( Q \) is a constant \cite{redyuk_jlt_2020, barreiro_preprint_2024}.
In Figure \ref{fig3}a, the region containing coefficients that satisfy \( |mn| < 300 \) is delineated by a solid black line.
The second method is based on selecting coefficients according to the condition \( |C_{m,n}| > Q \).
Although this approach initially requires a more complex computation to find the full set of coefficients, it achieves a lower MSE with the same number of coefficients as the first method.
Figure \ref{fig3}b illustrates the distribution of computed coefficients for the condition 
\( 10\lg(|C_{m,n}|/|C_{0,0}|) > -23.1 \)~dB.

\subsection{Convolutional Neural Network}
As previously described, in the first stage we use a neural network consisting of a single convolutional layer (see Figure \ref{fig_}, block 1). The feature matrix \( \mathbf{F} \) is structured with dimensions \( K \times N \), where \( K \) represents the number of symbols in the set and \( N \) denotes the number of features. The convolution kernel, denoted as \( \mathbf{x}_{CNN} \), has dimensions \( 1 \times N \). To train the kernel coefficients, we minimize MSE between the transmitted symbols \( \mathbf{a} \) and the equalized symbols \( \hat{\mathbf{b}} \), where \( \hat{\mathbf{b}} \) is the model's output for the set \( \mathbf{b} \). This training process adjusts the coefficients to effectively reduce the error and improve the accuracy of symbol recovery.

\begin{figure}[htbp]
\centering
\includegraphics[width=2.8in]{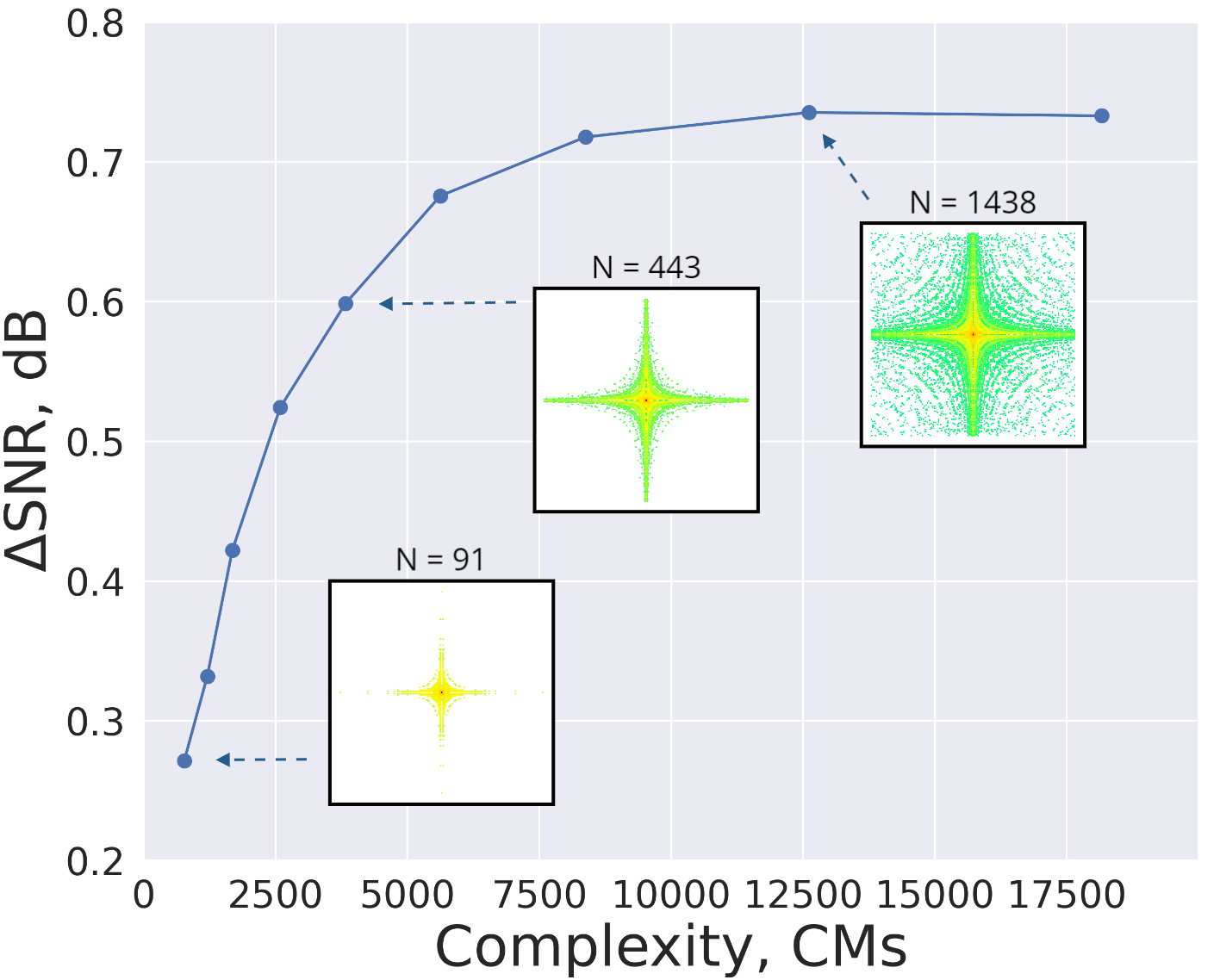}
\caption{The dependence of SNR gain on the computational complexity of the algorithm, measured in number of complex multiplications, for \( M=100 \).}
\label{fig4}
\end{figure}

Figure \ref{fig4} shows the dependence of the SNR gain on the computational complexity of the method for $M=100$. This was achieved by varying the parameter $Q$ when selecting coefficients according to the relation $|C_{m,n}|>Q$ while considering three symmetries. It can be observed that as the number of coefficients increases, the gain also increases, reaching up to 0.75~dB with an overall nonlinearity penalty of 2.2~dB. The saturation of the curve at computational complexity greater than 8000 multiplications indicates the inclusion of terms in equation (\ref{pbm}) with small perturbation coefficients that do not significantly contribute to the performance. The insets schematically illustrate the sets of coefficients corresponding to SNR gains of 0.27~dB, 0.6~dB, and 0.74~dB, as well as the number of unique coefficients $C_{m,n}$ in equation (\ref{pbm}).

The use of MSE as the loss function is justified by its differentiability, facilitating the training process. However, the primary quality metric for optical communication systems is BER, and using MSE raises several concerns. Firstly, MSE consider all symbols, including those decoded correctly, which do not contribute to BER. Secondly, in multi-level QAM modulation formats, the largest distortions due to nonlinearity affect symbols with the highest amplitudes, thereby significantly influencing MSE. However, since these symbols are only surrounded by neighboring symbols on two sides, their contribution to erroneous bits is not necessarily greater than others when decoded with a hard decision approach.

\subsection{Particle Swarm Optimization}
The main challenge of using BER as the objective function lies in the difficulty of computing its gradient. However, there exists a class of stochastic optimization methods, such as particle swarm optimization, genetic algorithms, and simulated annealing, which do not require the differentiability of the objective function. These methods can effectively optimize fitness function by exploring the solution space without relying on gradient information.

The challenge in applying these methods to minimize BER arises from the large number of optimized parameters \( C_{m,n} \). The hypothesis driving this study is that the set of parameters \( C_{m,n} \) that minimizes the MSE metric is closely approximates the set that minimizes BER. Under this assumption, the coefficients \( C_{m,n} \) derived from the CNN can serve as an initial approximation for stochastic optimization method  aimed at minimizing BER, thereby facilitating the search for its global minimum.
By using the CNN-derived coefficients \( C_{m,n} \) as a starting point for BER-based probabilistic methods, this approach has the potential to streamline the optimization process despite the non-differentiable nature of BER. This strategy represents a promising approach for effectively reducing BER in optical communication systems.

As a stochastic optimization method, we chose the particle swarm optimization \cite{wang_sc_2018}, where \( \text{BER}:\mathbb{R}^N\rightarrow\mathbb{R} \) represents the fitness function (see Figure \ref{fig_}, block 2). In this context, \( \mathbf{X}^i \in \mathbb{R}^N \) denotes the set of particles (\( i = 1...s \)), \( \mathbf{V}^i \in \mathbb{R}^N \) denotes the velocity of particle \( i \), \( N \) is the number of coefficients in \( \mathbf{x}_{CNN}  \), and \( s \) is the number of particles.
The function \( \text{Rand}(\mathbf{b}_l, \mathbf{b}_r) \) represents a multidimensional uniform distribution, where \( \mathbf{b}_l \) and \( \mathbf{b}_r \) are the lower and upper bounds of the solution space, respectively.
The boundaries are defined as \( \mathbf{b}_l = (1-\alpha)\mathbf{x}_{CNN} \) and \( \mathbf{b}_r = (1+\alpha)\mathbf{x}_{CNN} \), where \( \alpha \) is an optimized coefficient.
If a particle exceeds these boundaries in any component, that component's value is set to the corresponding boundary value.
The coefficients \( \omega \), \( \phi_p \), and \( \phi_g \) are constants, where \( \omega \) is the inertia weight, \( \phi_p \) and \( \phi_g \) are acceleration coefficients that determine the importance of the particle's best position and the swarm's best position, respectively.
The function \( \text{Rand}() \) denotes a vector whose components are independent random numbers uniformly distributed in the interval \( [0,1) \), chosen independently for each particle \( i \) at each iteration.
The outcome of this algorithm is a particle that achieves the minimum value of the fitness function (see Algorithm \ref{alg:alg1}).

\begin{algorithm}[htbp]
\caption{Particle Swarm Optimization}\label{alg:alg1}
\begin{algorithmic}
\STATE Initialize parameters: $s$, $b_l$, $b_r$, $max\_iter$, $\omega$, $\phi_p$, $\phi_g$
\STATE Initialize particles:
\FOR{$i \gets 1$ to $s$}
    \STATE Initialize position: $\mathbf{X}^i = \text{Rand}(\mathbf{b}_l, \mathbf{b}_r)$
    \STATE Initialize velocity: $\mathbf{V}^i = \text{Rand}(-(\mathbf{b}_r - \mathbf{b}_l), (\mathbf{b}_r - \mathbf{b}_l))$
    \STATE Initialize personal best: $\mathbf{P}^i = \mathbf{X}^i$
\ENDFOR
\STATE Initialize global best: $\mathbf{g} = \arg\min(\text{BER}(\mathbf{P}^i))$
\FOR{$t \gets 1$ to $max\_iter$}
    \FOR{$i \gets 1$ to $s$}
        \STATE Update velocity: $\mathbf{V}^i = \omega \cdot \mathbf{V}^i + \phi_p \cdot \text{Rand}() \cdot (\mathbf{P}^i - \mathbf{X}^i) + \phi_g \cdot \text{Rand}() \cdot (\mathbf{g} - \mathbf{X}^i)$
        \STATE Update position: $\mathbf{X}^i = \mathbf{X}^i + \mathbf{V}^i$
        \FOR{$d \gets 1$ to $N$} 
            \STATE $ \textbf{if } \mathbf{X}^i[d] < b_l[d] \textbf{ then } \mathbf{X}^i[d] = b_l[d]$
            \STATE $ \textbf{if } \mathbf{X}^i[d] > b_r[d] \textbf{ then } \mathbf{X}^i[d] = b_r[d]$
        \ENDFOR
        \IF{$\text{BER}(\mathbf{X}^i) < \text{BER}(\mathbf{P}^i)$}
            \STATE Update personal best: $\mathbf{P}^i = \mathbf{X}^i$
        \ENDIF
    \ENDFOR
    \STATE Update global best: $\mathbf{g} = \arg\min(\text{BER}(\mathbf{P}^i))$
\ENDFOR
\STATE \textbf{return} $\mathbf{g}$ as the found optimal solution
\end{algorithmic}
\label{alg1}
\end{algorithm}

\subsection{Results and Discussion}
\begin{figure}[htbp]
\centering
\includegraphics[width=3.45in]{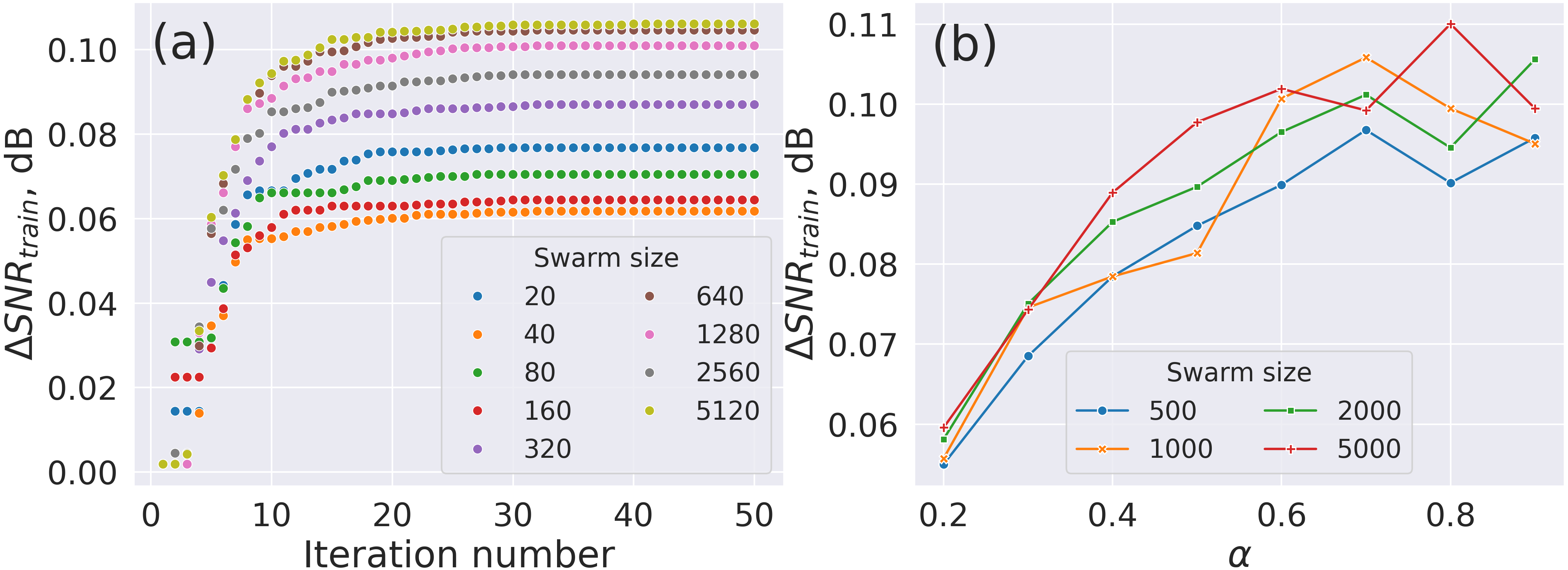}
\caption{The dependence of SNR gain on the number of iterations (a) and the range of solution space (b) when applying the PSO for different swarm sizes.}
\label{fig5}
\end{figure}

Figure \ref{fig5}a illustrates the dependency of the SNR gain after PSO relative to CNN as a function of the number of iterations for different swarm sizes.
It can be observed that regardless of the number of particles, the gain plateaus after 20 iterations.
Based on these results, we consistently used 30 iterations in subsequent experiments.

Figure \ref{fig5}b presents the dependence of the SNR gain on the parameter \(\alpha\), which defines the bounds of the solution space for PSO. It can be seen that an optimal value of \(\alpha\) exists in the range of 0.7 to 0.9, irrespective of the number of particles. For small values of \(\alpha\), the solution space becomes too compact, limiting the freedom to adjust \(C_{m,n}\) towards the global minimum of BER. Conversely, larger values of \(\alpha\) lead to an overly broad solution space, complicating convergence to the minimum. In subsequent experiments, \(\alpha\) was set to 0.9.

\begin{figure}[htbp]
\centering
\includegraphics[width=3.4in]{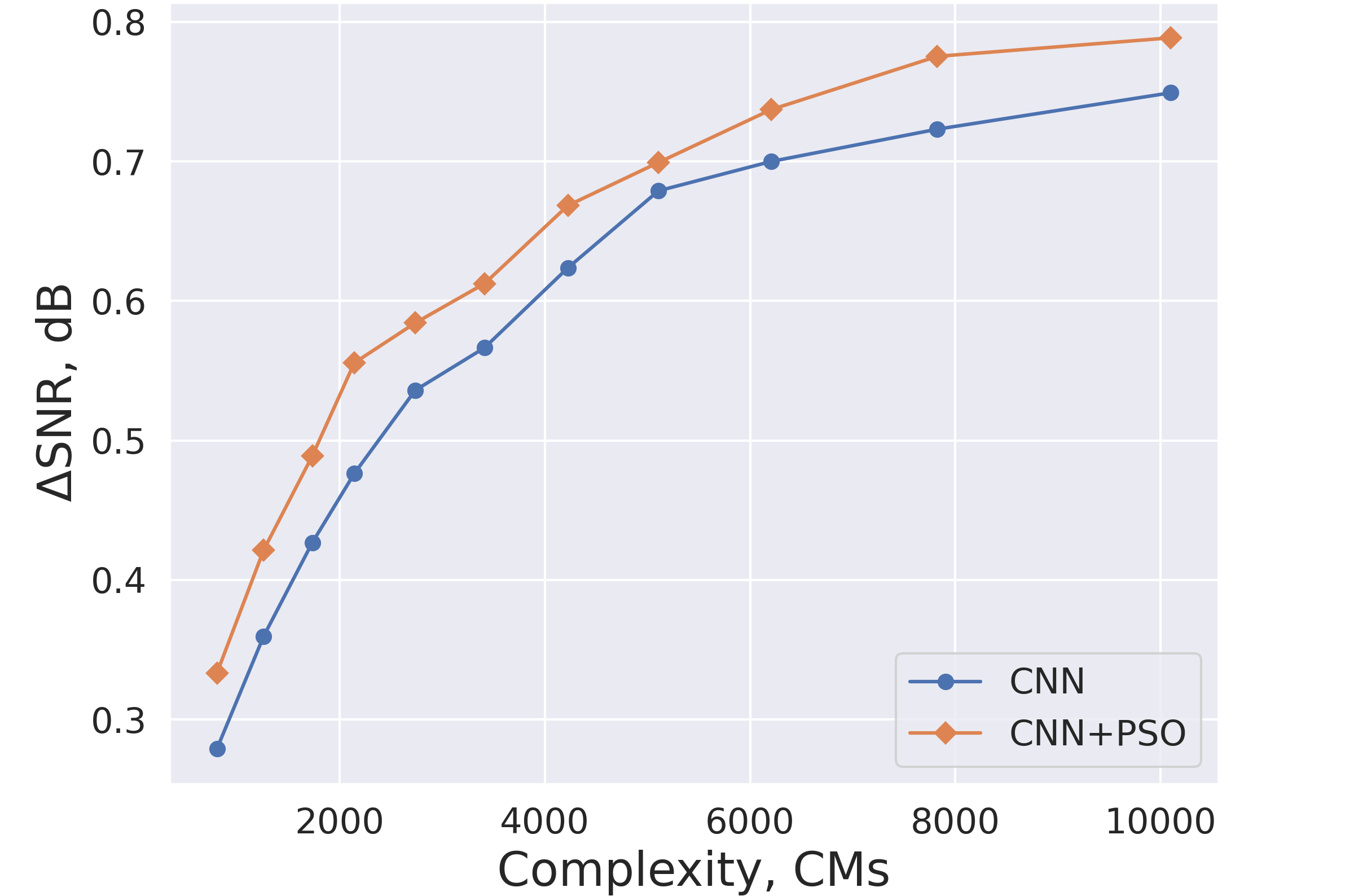}
\caption{Dependency of SNR gain on computational complexity, measured in number of complex multiplications, for CNN and CNN+PSO scenarios.}
\label{fig6}
\end{figure}

Figure \ref{fig6} illustrates the dependency of the SNR gain on the computational complexity after applying both CNN and PSO using 5000 particles in the swarm, 30 iterations, and \(\alpha=0.9\). The channel memory parameter is \(M=140\), and computational complexity varied by adjusting the parameter \(Q\) for selecting coefficients based on \(|C_{mn}| > Q\), while considering three symmetries. It can be seen that in all cases, there is an additional SNR gain after applying the two-stage method, ranging from 0.02 to 0.08~dB.

\begin{figure*}[!t]
\centering
\includegraphics[width=6.8in]{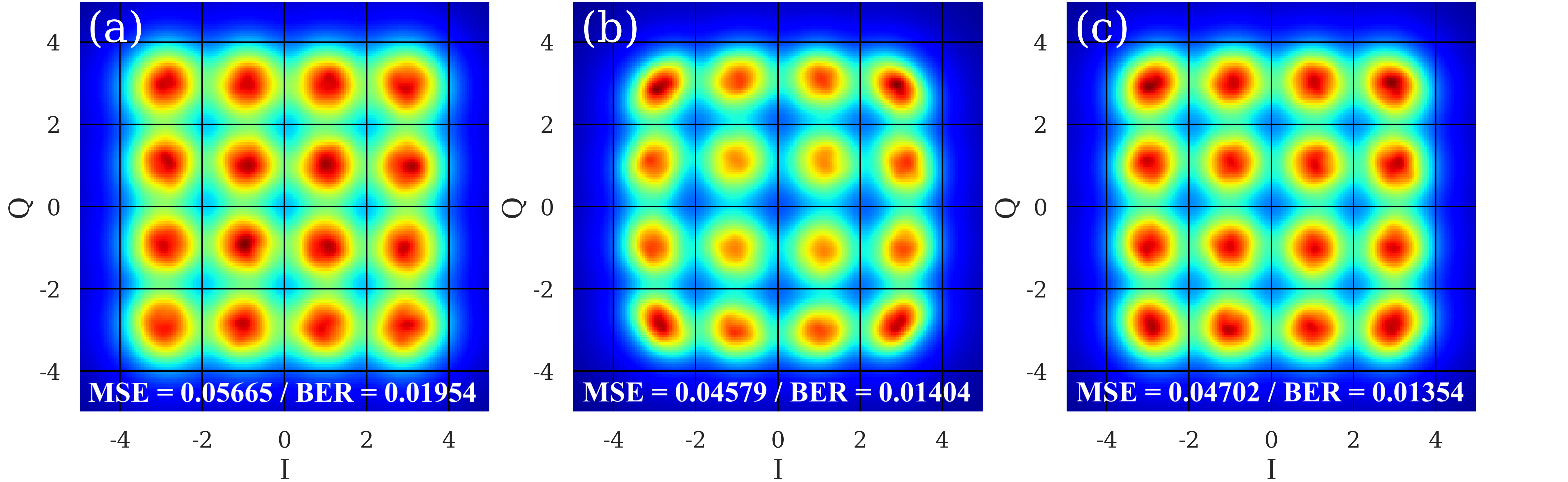}
\caption{Signal constellations: (a) before nonlinear distortion compensation, (b) after applying the model with perturbation coefficients computed using CNN, and (c) after applying the model with perturbation coefficients adjusted using PSO. Insets show the values of metrics MSE and BER.}
\label{fig7}
\end{figure*}

In Figure \ref{fig7}, signal constellations are depicted before nonlinear distortion compensation (a), after applying the compensation algorithm with coefficients computed using CNN (b) and after further refinement using PSO (c). Each subfigure includes corresponding metrics for MSE and BER.
After applying CNN (Figure \ref{fig7}b), the signal constellation exhibits a smoother and more compact distribution with improved separation between symbol clusters. Both MSE and BER metrics show significant reductions. However, it becomes apparent that the distribution of symbol clusters, particularly those with the highest amplitudes, becomes stretched and deviates from normal distribution. This deviation could potentially challenge subsequent forward error correction stages, as predictability and performance tend to degrade when signal distributions deviate from normal.

Following coefficient adjustment using PSO (Figure \ref{fig7}c), the distributions of the signal constellation again noticeably approximate normal distribution. Meanwhile, there is a slight increase in MSE, accompanied by a further decrease in BER.
Thus, the second stage of the scheme, involving coefficient refinement via PSO, significantly enhances the distribution of the signal constellation and further reduces the bit error rate. Both of these factors are crucial for enhancing the efficiency of the subsequent error correction process.

\section{Computing Perturbation Coefficients Without Direct Knowledge of Transmitted Symbols}
In the experiments above, we assumed that the transmitted symbols $\mathbf{a}$ was known and we used these symbols as target. However, in real systems, nonlinear equalizers can only utilize the received symbols $\mathbf{b}$.
We propose and investigate a method for determining perturbation coefficients without the knowledge of the symbols $\mathbf{a}$. 
The idea of this method involves replacing $\mathbf{a}$ in the previously described approach with the symbols obtained after demodulating $\mathbf{b}$ using hard decision decoding.
Given that the BER of $\mathbf{b}$ is 0.02, we can infer that the hard decision decoding of $\mathbf{b}$ contains approximately 98\% correct and 2\% erroneous symbols compared to the symbols~$\mathbf{a}$.

\begin{figure}[htbp]
\centering
\includegraphics[width=3.1in]{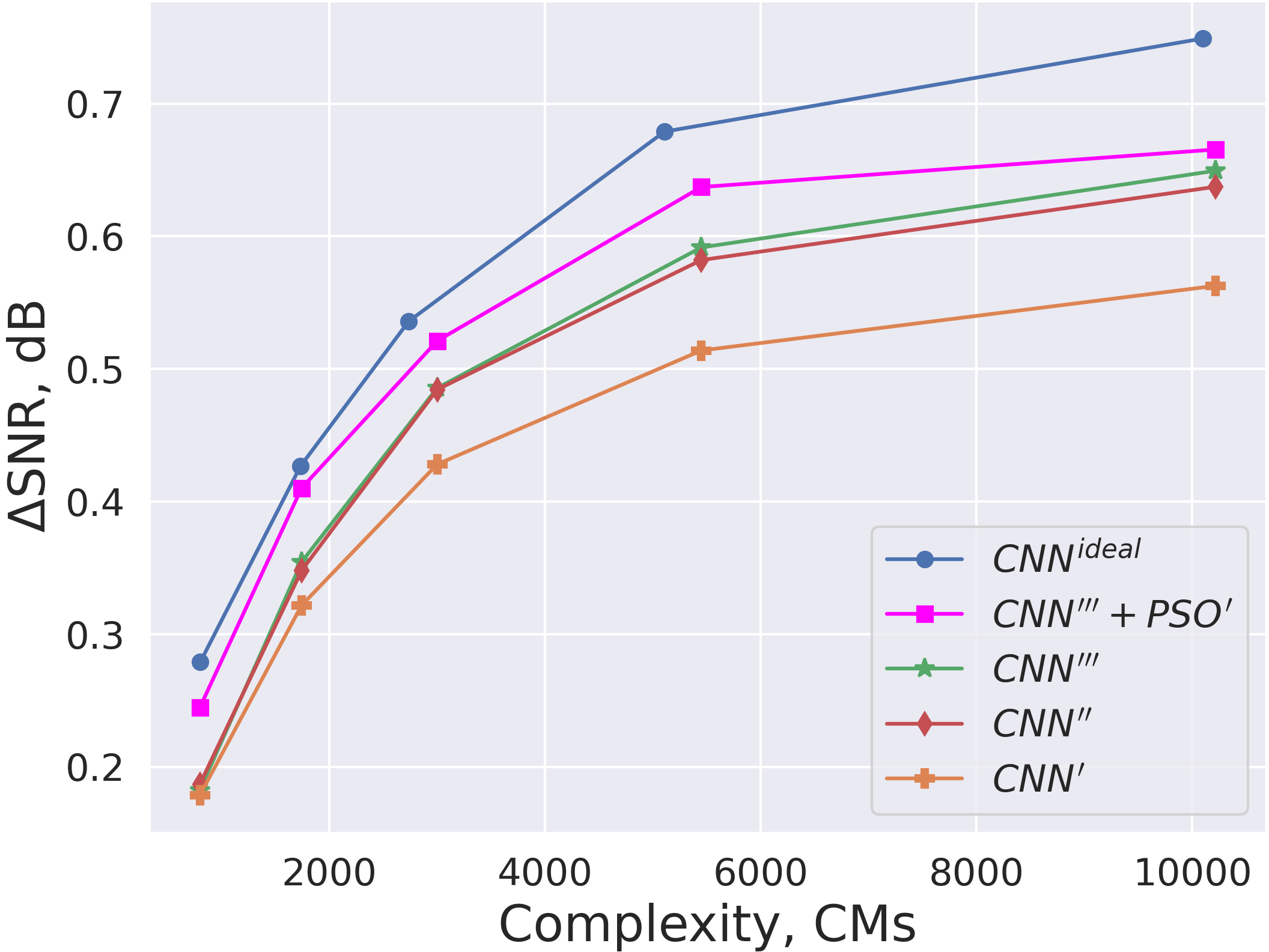}
\caption{Dependency of SNR gain on computational complexity in terms of complex multiplications for CNN only and CNN+PSO approaches.}
\label{fig8}
\end{figure}

Firstly, we applied this technique for CNN stage. For experiments, the channel memory parameter was set to $M = 140$, and the computational complexity was varied by adjusting the parameter $Q$, which selects coefficients according to the condition $|C_{m,n}| > Q$. Using proposed approach during training, an SNR improvement of approximately 0.55~dB can be achieved (Figure \ref{fig8}, orange curve $CNN^{I}$), which is significantly less than the improvement obtained when training with the known symbols $\mathbf{a}$. 
To enhance performance, we propose additional iterations of CNN training (see CNN-stage of Algorithm \ref{alg:alg2}).
In the $i$-th iteration, the target set consists of symbols $\mathbf{\hat a}^{(i)}=HD(\mathbf{\hat b}^{(i-1)})$, where $\mathbf{\hat b}^{(i-1)}$ is the result of applying the trained CNN model from the previous iteration $i-1$ to the symbols $\mathbf{b}$. Following this iterative approach, after three iterations an SNR improvement of approximately 0.65~dB was achieved (Figure \ref{fig8}, green curve $CNN^{III}$). No further SNR improvement is observed beyond three iterations.

The question arises on how to implement training at the PSO stage in this context, as optimizing BER without knowing the symbols $\mathbf{a}$ is not feasible. If we similarly take the symbols $\mathbf{b}$ as the input and the symbols $\mathbf{\hat a}^{(4)}=HD(\mathbf{\hat b}^{(3)})$ as the target set for PSO stage, it is evident that no SNR improvement can be achieved relative to the last CNN iteration. Therefore, we propose adding random noise to the symbols in the set $\mathbf{b}$ before training at the PSO stage. The hypothesis is that this approach introduces random distortions that are challenging to compensate for, artificially increasing the BER of symbols $\mathbf{b}$. Consequently, the model during optimization will focus on compensating for deterministic nonlinear distortions rather than random ones. Finally, by applying the trained model to the noise-free symbols $\mathbf{b}$, we can achieve an additional SNR improvement.

Pink curve of Figure~\ref{fig8} shows the results of CNN+PSO scheme for size of swarm $s = 5000$ and $\alpha = 0.9$. It can be observed that across all cases within the considered region, there is an SNR improvement relative to CNN, with the magnitude ranging from $0.02$ to $0.06$~dB.

For clarity, the proposed approach is outlined in Algorithm \ref{alg2}. Here, $HD(\mathbf{b})$ denotes the hard decision decoding of symbols $\mathbf{b}$, $PBM(\mathbf{x}, \mathbf{b})$ represents the substitution of coefficients $\mathbf{x}$ and symbols $\mathbf{b}$ into equation (\ref{pbm}), $CNN(\mathbf{a}, \mathbf{b}) \xrightarrow[MSE]{learning} \mathbf{x}_{CNN}$ refers to training the CNN model using the target sequence $\mathbf{a}$, the training set $\mathbf{b}$, and the loss function $MSE$. Similarly, $PSO(\mathbf{a}, \mathbf{b}) \xrightarrow[BER]{optimization} \mathbf{x}_{PSO}$ describes the PSO optimization process using the $BER$ metric.

\begin{algorithm}[H]
\caption{HD-based CNN+PSO}\label{alg:alg2}
\begin{algorithmic}
    \STATE $\mathbf{\hat b}^{(0)} = \mathbf{b}$
    \FOR{$i \gets 1 \text{ to } 3$}
        \STATE $\mathbf{\hat a}^{(i)} = HD(\mathbf{\hat b}^{(i-1)})$
        \STATE $CNN(\mathbf{\hat a}^{(i)}, \mathbf{b}) \xrightarrow[MSE]{learning} \mathbf{x}^{(i)}_{CNN}$
        \STATE $\mathbf{\hat b}^{(i)} = PBM(\mathbf{x}^{(i)}_{CNN}, \mathbf{b})$
    \ENDFOR
    \STATE $\mathbf{\hat a}^{(4)} = HD(\mathbf{\hat b}^{(3)})$
    \STATE $PSO(\mathbf{\hat a}^{(4)}, \mathbf{b_{noise}}) \xrightarrow[BER]{optimization} \mathbf{x}^{(1)}_{PSO}$
    \STATE $\mathbf{\hat b}^{(4)} = PBM(\mathbf{x}^{(1)}_{PSO}, \mathbf{b})$
\end{algorithmic}
\label{alg2}
\end{algorithm}


\section{Conclusions}
This study introduced an enhanced approach for compensating fiber nonlinearity based on perturbation theory. The method employs a two-stage scheme to effectively compute perturbation coefficients.
In the first stage, convolutional neural networks were utilized to estimate coefficients by minimizing the mean squared error between equalized and transmitted symbols. Subsequently, in the second stage, the particle swarm optimization was employed to further refine these coefficients, aiming to minimize the bit error rate. Application of this approach on a 2000~km communication link using a 16QAM signal demonstrated significant improvements in signal-to-noise ratio of up to 0.8~dB. This enhancement was accompanied by improvements in the distribution of the signal constellation.
Moreover, the proposed method for computing perturbation coefficients without explicit knowledge of transmitted symbols resulted in an SNR improvement of up to 0.66~dB. This approach shows promise for enhancing the performance of optical communication systems, particularly in mitigating nonlinear distortions.

\section{Acknowledgment}
This work was supported by the Analytical Center for the Government of the Russian Federation in accordance with the subsidy agreement (identifier 000000D730324P540002) and the agreement with the Novosibirsk State University dated December 27, 2023 No. 70-2023-001318.

\end{document}